\documentclass[12pt]{iopart}
\usepackage{graphicx}
\usepackage{url}
\newcommand{\bstar}{\beta^{*}}
\newcommand{\emittn}{\epsilon_{n}}
\newcommand{\emittl}{\epsilon_{l}}
\newcommand{\positron}{\mathrm{e}^{+}}
\newcommand{\electron}{\mathrm{e}^{-}}
\newcommand{\epem}{\positron\electron}
\newcommand{\neutron}{\mathrm{n}}
\newcommand{\PbAZ}{^{208}\mathrm{Pb}^{82+}}
\newcommand{\PbBFPP}{^{208}\mathrm{Pb}^{81+}}
\newcommand{\PbEMD}{^{207}\mathrm{Pb}^{82+}}
\newcommand{\lumiunits}{\mathrm{cm}^{-2}\,\mathrm{s}^{-1}}
\newcommand{\LNN}{L_{\mathrm{NN}}}

\begin{document}

\title[LHC as nucleus-nucleus collider]{The LHC as a Nucleus-Nucleus
Collider}

\author{John M. Jowett}

\address{Accelerators and Beams Department, CERN, CH-1211 Geneva 23, Switzerland}
\ead{John.Jowett@cern.ch}

\begin{abstract}

This paper begins with a summary of the status of the Large Hadron
Collider at CERN, including the lead-ion injector chain and the plans
for the first phases of commissioning and operation with colliding
proton beams.  In a later phase, the LHC will collide lead nuclei at
centre-of-mass energies of 5.5~TeV per colliding nucleon pair.  This
leap to 28~times beyond what is presently accessible will open up a new
regime, not only in the experimental study of nuclear matter, but also
in the beam physics of hadron colliders.  Ultraperipheral and hadronic
interactions of highly-charged beam nuclei will cause beam losses that
dominate the luminosity decay and may quench superconducting magnets,
setting upper limits on luminosity and stored beam current.  Lower
limits are set by beam instrumentation.  On the other hand, coherent
radiation by the nuclear charges should provide natural cooling to
overcome intra-beam scattering.  As with protons, a flexible, staged
approach to full performance will test the limits and make optimal use
of scheduled beam time.

\end{abstract}


\submitto{\JPG}

\maketitle

\section{Introduction}

CERN and its world-wide network of collaborating institutions are almost
at the end of the long road from the first public ``Feasibility Study of
a Large Hadron Collider in the LEP Tunnel'' in 1984~\cite{Asner:1984jv}
to colliding protons and heavy (lead) nuclei in the LHC.
A recent survey of  the machine design, superconducting technology and
the accelerator
physics relevant to its performance as a proton collider can be found
in~\cite{Evans2007}.
After a brief update on the status (as of February 2008), this paper
will describe the LHC's second role as an
ultra-relativistic nucleus-nucleus collider,
with emphasis on performance expectations and the physical
phenomena that will limit luminosity.
The heavy-ion physics programme itself
is described in other papers at this conference.

The project has required a host of technological
developments~\cite{Evans2007,lhc1} of which industrial scale production
of the 15~m long dipole magnet is the most prominent.  There will be
1232 of these, operating with their superconducting coils bathed in
superfluid helium at a temperature of 1.9~K.

Inevitably, in such a large and immensely complex enterprise there have
been a few setbacks.  Well-known examples were difficulties with the
main cryogenic helium transport line, the low-$\beta$ triplet
quadrupoles and the interconnect modules that have to keep the beam
impedance low while compensating the thermal expansion of sections of
the beam screen (inside the vacuum chamber) during warm-up and cool-down
of the machine.  As part of the enormous efforts made in recent years to
minimise slippage of the schedule, technical solutions have been found
and implemented.  Installation of the collider's hardware is now
complete and hardware, then beam, commissioning will soon be under way.

\subsection{Luminosities}

The nucleon-nucleon luminosity, $\LNN$, and ion luminosity, $L$, of the
LHC are given by
\begin{equation}\label{eq:lumidef}
\LNN = A^2 L =
\frac{{A^2 N_b^2 k_b f_0}}
{{4\pi \sigma _x \sigma _y }}F(\theta_c,\sigma^*,\sigma_z )
= \frac{{A^2 N_b^2 k_b f_0 \gamma }}
{{4\pi \, \emittn \beta ^* }}F(\theta _c,\sigma^*,\sigma_z )
\end{equation}
where
$f_0$ is the revolution frequency,
$N_b$ is number of particles (protons or ions) per bunch,
$k_b$ is the number of bunches per beam,
$\gamma=E/(m c^2)$ is the usual relativistic factor;
$\emittn=\sqrt{\gamma^2-1} \,\sigma_{x,y}^2 /\beta^*$
is the
``normalised'' (independent of beam momentum $p$) emittance
related to the beam size $\sigma^*$ and
$\beta^*$, the optical function at the interaction point (IP)
(the beams are round so these quantities are the same in both planes);
finally,
$F(\theta_c,\sigma^*,\sigma_z ) = (1 + ( \theta _c \sigma _z/2\sigma ^*
)^2 )^{-1}$
is a  reduction factor from the half-crossing angle, $\theta _c$,
and bunch length $\sigma _z$.

With its nominal bending field of 8.3~T,
the LHC will provide collisions at the centre-of-mass energy
\begin{equation}  \label{eq:Enominal}
  \sqrt{s}=
   \left\{
   \begin{array}{ll}
         14  \, \mathrm{TeV}                        &\mbox{(p-p)}\\
         1.15\, \mathrm{PeV}=5.52\, A\, \mathrm{TeV}
&\mbox{($\PbAZ$--$\PbAZ$)}
   \end{array}
   \right.
 \end{equation}
in the first years of operation, eventually aiming for the design
nucleon-nucleon luminosities
 \begin{equation}\label{eq:lumiDesign}
   L_{\mathrm{NN}}  \approx
   \left\{
   \begin{array}{ll}
         10^{34}\, \lumiunits                &\mbox{(p-p)}\\
         4.3\times 10^{31}\, \lumiunits\
         =  10^{27} A^2 \lumiunits           &\mbox{($\PbAZ$--$\PbAZ$)}
   \end{array}
   \right.
\end{equation}
Each of the  four large experiments  will study p-p
collisions while heavy-ion (nuclear) collisions will be provided to
ALICE, ATLAS and CMS.

\section{Present status of the LHC}

\subsection{Schedule}
At present, we expect the whole machine to be cold
by early June 2008, some 2-3 weeks behind the schedule published
in October 2007.
The  schedule, which is constantly updated at~\cite{LHChomepage},
is technically feasible but
remains sensitive to any major
new problem.

Proton commissioning and operation for physics
will continue through 2008 and 2009.
At present, the target date for the first Pb-Pb collisions is
at the end of the 2009 run, before the winter shutdown.
The ion injectors will not run in 2008 and must be made ready earlier in
2009.

\subsection{Commissioning with proton beams}

Since the injector chain and transfer lines from the SPS
to the LHC are already fully capable of delivering the required
proton beams,
commissioning of the accelerator with protons will start with
procedures to achieve injection, RF capture and
good lifetime at injection energy for
single, moderate intensity bunches.  Thereafter
commissioning will proceed in defined stages (labelled A--D), gradually
increasing $k_b$, $N_b$
and ``squeezing'' $\beta^*$ to smaller values.
Further details of the beam parameters and procedures are given under
``Commissioning'' at~\cite{LHChomepage}.

Experience of previous colliders shows that it is hard to predict the
time necessary to achieve first collisions although the estimate for the
LHC is about 2~months.  This machine will require particular care to
establish a precise knowledge of the orbit and beam optics.  The
aperture available to the beam is small and the stored energy in the
beams will be quite unprecedented.  Therefore the luminosity attainable
will depend on the ability to protect the machine from losses ($\sim
1/\beta^*$), experience with the collimation system and other factors.
Unlike any previous collider, the collimation system's primary purpose
is to clean the beam and protect the machine rather than to reduce
backgrounds in the experiments.

\section{Ion Injector Chain}

The heavy ion beams required for the LHC are much more demanding in
intensity and emittance than those used in the SPS fixed target
programme.
This has required a new electron-cyclotron resonance (ECR)
ion source, the electron cooling ring LEIR and many other changes and
upgrades  to the CERN injector complex~\cite{lhc3}.  Together these
constituted the bulk of the cost of the ``Ions for LHC'' project.

Two reference sets of LHC beam parameters
(dubbed ``Early'' and ``Nominal'')
correspond to different modes of operation  of  the injectors
(see \cite{lhc1,lhc3} and Table~\ref{tab:parameterlist}).

The status of the injector chain was most recently
reviewed in~\cite{APAC2007ions};
only more recent developments are summarised in the following.

\begin{description}

\item[Source and Linac3] achieved adequate   intensity  for
Early beam
(record of 31~e\,$\mu$A of Pb$^{54+}$ out of the linac).
The stability and reliability required for Nominal beam will be supplied
by an upgrade of the source generator to~18 GHz.
Numerous other improvements have been implemented or are on the way.

\item[LEIR] is working well for the Early beam and there has been
progress towards Nominal.

\item[PS and transfer lines] will require further work for the Nominal
beam.

\item[SPS] was commissioned for ions in late 2007 although there were
substantial delays with some hardware.  The Early beam parameters were
essentially achieved but there are concerns about beam losses on the
longer injection plateau needed for Nominal.  If these are not reduced
by further development on the RF system, it may be necessary to change
the LHC filling scheme to shorten the plateau.  Nevertheless a $\PbAZ$
beam was ejected along one of the transfer lines from the SPS towards
the LHC.

\end{description}

\section{Pb-Pb collisions}\label{sec:PbPbCollisions}

Historically, the Nominal parameters (Table~\ref{tab:parameterlist})
were defined, on the basis of experimental requirements, many years ago.
The ion injector chain was accordingly
designed to provide appropriate beam intensities.
More recently, it was recognised that
the Pb-Pb luminosity in the LHC might be limited by new beam physics
effects, not seen in any previous collider.
A peak luminosity
$L_0\simeq10^{27}\mathrm{cm}^{-2}\mathrm{s}^{-1} $ has been kept as a
goal although quantitative uncertainties in the performance limits
(discussed below) might limit it to values 2--3~times less.

The Early parameters (Table~\ref{tab:parameterlist})
were introduced more recently~\cite{lhc1}
as a first step in a staged commissioning plan, allowing more rapid
commissioning of the injectors, exploration of the new performance
limits in the LHC itself, and a luminosity sufficient for initial
physics.

\begin{table}
   \begin{center}
   \begin{tabular}{|l|c|c|c|}
      \hline
      Parameter      & Units & Nominal  & Early   \\
      \hline
      Energy/nucleon      & TeV & \multicolumn{2}{c|}{2.76}    \\
      Peak luminosity $L_0$ & $ \mathrm{cm}^{-2}\mathrm{s}^{-1}$ & $\sim 10^{27} $ & $\sim 5.\times 10^{25}$  \\
      No. of bunches $k_b$  &  & 592  & 62  \\
      Bunch spacing  & ns & 99.8  & 1350  \\
      Optics ($\bstar$) at IP2/IP1,5  & m & 0.5/0.55 & 1.0  \\
      No. of Pb ions/bunch $N_b$      & &  \multicolumn{2}{c|}{$7.\times 10^{7}$ }     \\
      Transverse normalised RMS emittance $\emittn$      & $\mu\mathrm{m}$ & \multicolumn{2}{c|}{1.5 }     \\
      Longitudinal emittance/charge $\emittl$      & $ \mathrm{eV\,s}$ & \multicolumn{2}{c|}{2.5 }     \\
      Luminosity half-life (1,2,3 experiments)      & h & 8, 4.5, 3 &14, 7.5, 5.5 \\
%
%            & &  &   \\
%            &  & \multicolumn{2}{c|}{ }     \\
%
     \hline
   \end{tabular}
   \end{center}
   \caption{Selected performance parameters for the ``Nominal'' and ``Early''
            Pb-Pb collision modes; for full details see chapter~21  		of~\cite{lhc1}.
           }\label{tab:parameterlist}
\end{table}

Ultraperipheral and hadronic interactions of beam nuclei with other
nuclei, either in the opposing colliding beam or in the stationary beam
environment are at the root of the main performance limits of the LHC
when it collides heavy nuclei.

\subsection{Ultraperipheral collision processes}

The cross section for  free  $\epem$ pair-production
%% \[
%%      Z_1  + Z_2  \to Z_1  + \electron  + \positron  + Z_2
%% \]
in  collisions of nuclei with charges $Z_1$ and $Z_2$ is
%%  \begin{equation}\label{eq:sigmaPP}
%%  \sigma _{\mathrm{PP}}  = \frac{Z_1^2 Z_2^2 \alpha^2 r_e^2 }{\pi }
%%  \left[
%%        \frac{224}{27}
%%        \log\left( 2\gamma _{\mathrm{CM}} \right)^3  +  \cdots
%%  \right]
%%   \approx   2.\times 10^4\, \mbox{b}.
%%  \end{equation}
%
$ \sigma _{\mathrm{PP}} \propto Z_1^2 Z_2^2   \approx   2.\times 10^4\, \mbox{b}$
for Pb-Pb at the LHC.
A small fraction of the pairs are produced with the electron
bound to one  nucleus in \emph{bound-free pair production} (BFPP):
\begin{equation}\label{eq:BFPP}
Z_1  + Z_2  \to \left( {Z_1  + \electron } \right)_{{\mathrm{1s}}_{{\mathrm{1/2}}} , \ldots }  + \positron  + Z_2
\end{equation}
and with a  cross section~\cite{MeierBFPP}
depending much more strongly on  $Z$:
\begin{equation}\label{eq:sigmaBFPP}
\sigma _{\mathrm{BFPP}}
       \simeq
       Z_1 ^5 Z_2 ^2
       \left[ {A\log \gamma _{CM}  + B} \right]
  \approx
  \mbox{281 b} ,
\end{equation}
for Pb-Pb at the LHC; \cite{MeierBFPP} gives values for the constants
$A$ and $B$.
This process, together with the electromagnetic dissociation
(EMD) via the Giant Dipole Resonance
\begin{equation}\label{eq:EMD}
\PbAZ + \PbAZ  \longrightarrow \PbAZ  + \PbEMD + \neutron
\end{equation}
dominates the intensity loss (``burn-off'') from
collisions~\cite{BaltzBFPP,Jowe0}.
The collision products have a different  charge-to-mass ratio
and are lost from the main beam~\cite{KleinBFPP}.
The beam aperture and optics are such that the $\PbEMD$
nuclei produced by EMD are lost safely in the momentum collimation system.
However the beams of
$\PbBFPP$
ions from BFPP, emerging from each side of each collision point,
strike the beam screen inside one of the first
superconducting bending magnets at the start of  the main arc
(the dispersion suppressor section) of the LHC.
The 281~kHz loss rate at nominal luminosity generates
25~W of heating power in a $\sim 1\,\mathrm{m}$ long spot.

Detailed analysis~\cite{PAC05BFPP}, including simulations of the
hadronic showers, and revised estimates of the tolerable energy
deposition (thermodynamics of liquid He and heat transfer), suggest that
the magnets are not likely to quench because of BFPP beam losses;
however,
quenches remain possible within the uncertainties.  Additional beam loss
monitors have been installed around the IPs to monitor these losses in
LHC operation and strategies are being prepared to redistribute them to
some extent.

Despite much smaller rates during the RHIC Cu-Cu run, it was just
possible to detect this process~\cite{RHICBFPP} and test the methodology
used to predict the energy deposition in the magnet coils and signals on
beam loss monitors in the LHC.

\subsubsection{Collimation inefficiency}

The collimation system is essential to protect the LHC machine from
particles that would be lost causing magnet quenches or damage.

The principle of collimation for protons is that particles at large
amplitudes undergo multiple Coulomb scattering in a
sufficiently long, (carbon) primary collimator,
deviating their trajectories onto properly placed
secondary collimators which absorb them in hadronic showers.
However ions undergo nuclear fragmentation or EMD before scattering
enough  so the secondary collimators are
ineffective~\cite{EPAC04IonCollimation}.
The machine
then acts as a spectrometer with isotopes lost in other locations,
including superconducting magnets, with consequences as described above.
Simulation of these processes requires detailed nuclear physics input
with cross sections for many fragment channels.
Again, the results suggested the locations of additional beam monitors.

This may turn out to be a more severe limit on Pb-Pb luminosity than
BFPP.  Nevertheless it should be kept in mind that the conventional
(1996) quench limit (tolerable heat deposition in superconducting magnet
coils) now appears pessimistic.  This is also a soft limit:  losses are
evaluated with the hypothesis that the single-beam (not including
collisional) losses have reached a level corresponding to a lifetime of
12~min.  Meanwhile the simulations have been successfully benchmarked
with Pb beams and an LHC collimator in SPS.

Since there will be a Phase 2 Collimation upgrade for p-p operation, we
are looking at what might be included to improve collimation
efficiency for ion beams (cryogenic
collimators, crystals, magnetic collimation, optics changes, etc.)

\subsubsection{Beam Instrumentation}

\begin{figure}
\begin{center}
\includegraphics*[width=10cm]{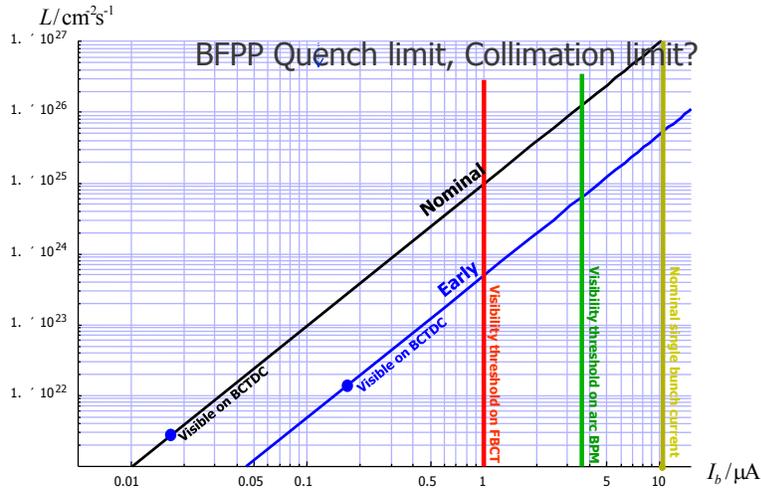}
\end{center}
\caption{Luminosity, $L$,
        vs. single bunch current $I_b$, limited from below by the dynamic range
        of BPMs and current monitors (BCTDC, FBCT) for both Early and Nominal bunch
        filling schemes. $L$ is limited by BFPP and total current $k_b I_b$ by
        collimation inefficiency.
        Operation is restricted to the diamond at upper right.
        }
\label{fig:OperationalParameterSpaceWithLeadIons}
\end{figure}

The total  charge in a Pb bunch is only a factor 3-4 above the  lower
limit of visibility
on the beam position monitors (BPMs); therefore it will always be
necessary
to inject close to nominal bunch current and (very likely) dump beams
when their intensity decays below this threshold.
There are similar limits on the beam current monitors, all indicated in
Figure~\ref{fig:OperationalParameterSpaceWithLeadIons}.   Any limits on
\emph{total} bunch will be respected by adjusting the number of bunches.

The methods for measuring beam sizes and emittances are also limited,
with greater reliance on beam-gas ionization monitors and Schottky
spectra than for protons.

\subsubsection{Beam and luminosity lifetime}

\begin{figure}
\begin{center}
\includegraphics*[width=14cm]{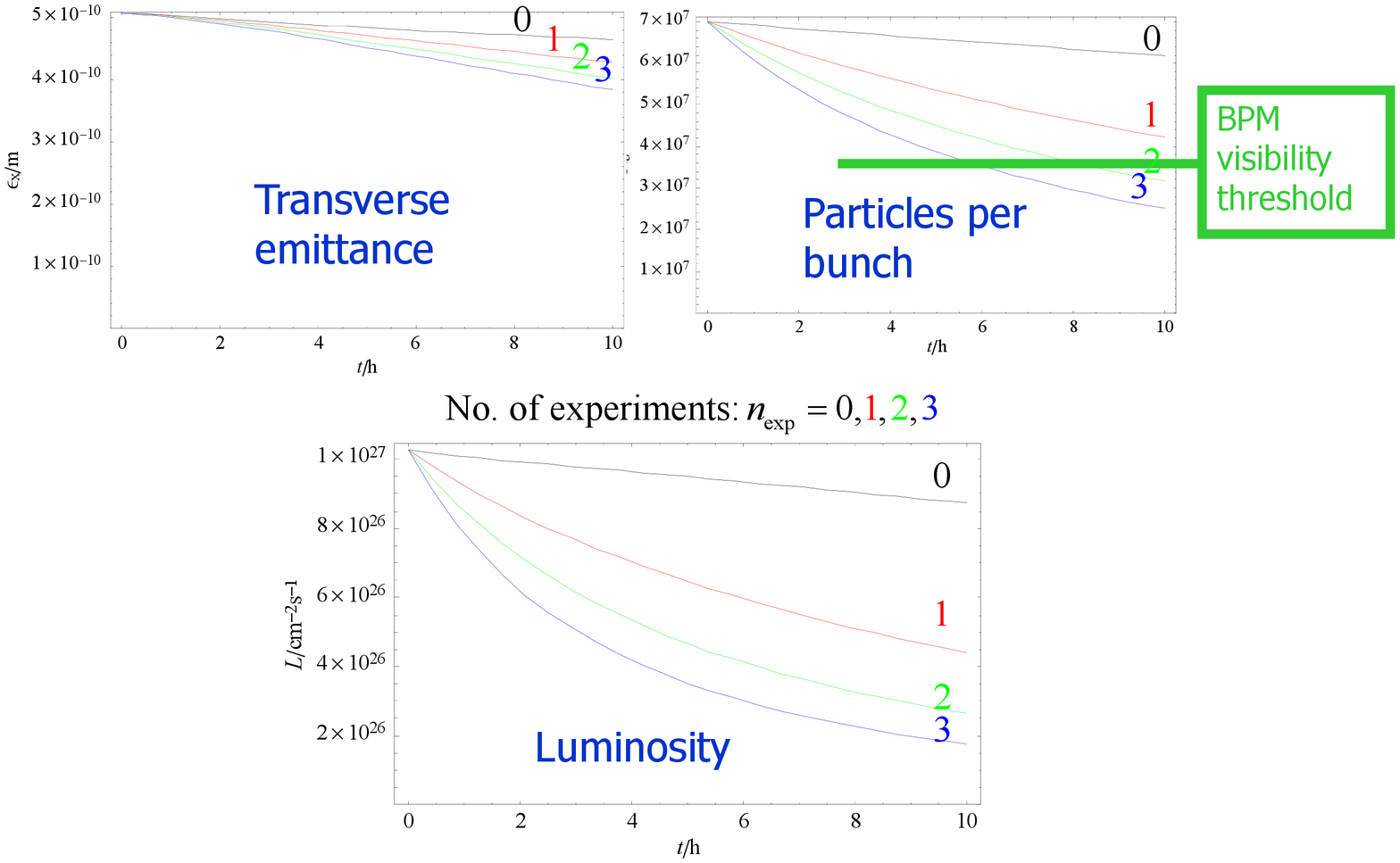}
\end{center}
\caption{Evolution of emittance, bunch intensity and luminosity
in an ``ideal''
fill, starting from design parameters giving Nominal luminosity. Cases
of 0,1,2 or 3 active experiments are shown.
Beams will most likely be dumped when the intensity decays to the BPM visibility threshold.
}
\label{fig:LuminosityEvolution}
\end{figure}

Once beams are put into collision, the subsequent evolution of the
intensity and lifetime depends on the interplay of a number of effects.
Predictions of the net results are shown in
Figure~\ref{fig:LuminosityEvolution}.

Beam-gas interactions, reducing intensity and increasing emittance, are
not expected to be significant once good vacuum conditions are
established.

Intra-beam scattering (IBS) or multiple Coulomb scattering within
bunches tends to blow up the beams on a time scale of several
hours.
However, since the nuclear charges radiate coherently at relevant
wavelengths, the LHC ions will be the first hadron beams to be
significantly affected by synchrotron radiation damping.
Somewhat surprisingly, radiation damping for Pb ions is
about twice as fast  as for protons (see Chapter~21 of~\cite{lhc1})
and fast enough to overcome IBS at full intensity, hence the shrinking
emittance in
Figure~\ref{fig:LuminosityEvolution}.  In addition (although it is not
immediately apparent in these plots), longitudinal
RF noise is also being used to counteract the damping of the
longitudinal emittance, keeping it roughly constant.  This helps to
reduce the effect of IBS on transverse emittances.

The intensity decay is dominated by the strong ``burn-off'' in
collisions from the large electromagnetic cross-sections.  Instantaneous
beam and luminosity lifetime are reduced in proportion to the number of
active experiments.  This is partly compensated by the faster emittance
damping as the intensity drops.  But beams must also be dumped sooner
and the average and time-integrated luminosity will depend strongly on
the
time taken to dump, recycle, refill, ramp and re-tune the machine for
collisions.  Clearly, the \emph{integrated luminosity per experiment
will fall as more experiments take collisions}.

\subsection{Commissioning Pb-Pb collisions}\label{sec:Pbcommissioning}

The ``hot-switch'' to Pb-Pb collisions will be done when the LHC is
already operational with protons and the ion injector chain is ready.
It will not be a start-up from shutdown.  The rapid commissioning
strategy is based on the principle of making the absolute minimum of
changes to the working p-p configuration.

The quasi-static \emph{magnetic fields} of the LHC magnets will have
\emph{exactly the same effect on the spatial trajectory} of a Pb ion as
on a proton (for equal momentum per charge, $p/Z$, or magnetic
rigidity).  Thus, all the magnetic settings established in p-p operation
for the transfer, injection, ramp, and squeeze of the ATLAS and CMS
collision optics should also work for Pb ions.  Moreover, the nominal
emittances are chosen to give equal beam sizes so all related
considerations (e.g., of aperture) should be similar.  The main change
to the magnetic cycle will be the completion of a $\beta$-squeeze for
ALICE; we expect that the experience gained by then with the other
experiments will allow this to be commissioned quite efficiently.
Indeed, most or all of this setup may well have been done with proton
beams.

The LHC beams see few externally applied electric fields apart from the
time-dependent electric fields of the RF system, localised near Point~4.
Adjustments of the RF frequency and phase will compensate for the change
in speed and revolution frequency of the ions---``energy matching'' and
``capture'' at injection, then a calculable shift at each energy in the
ramp---will match the ions' orbit to that of the protons.

Other operational differences will arise from the different bunch
filling patterns and adaptation of the beam instrumentation but these
should generate little overhead.

This may seem inconsistent with the experience at RHIC, which has
switched species several times, typically from A-A to p-p.
However RHIC requires more complicated  changes to the magnetic fields
(transition crossing in the ion ramp, polarized proton beams) that are
not necessary in the LHC.
A better comparison might be with the CERN ISR
which switched from p-p to light ion collisions
a few times in the  late 1970s.
Those switches went very quickly~\cite{Myers:1977zi},
in less than a day, precisely
because the machine was kept magnetically identical.

After a first run with the Early beam, we will gradually push up the
number of bunches towards Nominal, always maximising the single-bunch current
within the overall limitations.
As with protons, it may be worth changing the bunch filling pattern
in the light of better quantitative knowledge of the performance limits.

\section{Beyond Pb-Pb collisions}\label{sec:beyond}

So far resources have been concentrated  on the ``baseline'' of p-p and
Pb-Pb collisions.
However the heavy-ion physics programme at the LHC is expected to include
further stages not yet scheduled within the CERN programme.
These  may include:

\begin{description}

\item[p-Pb collisions]  are a crucial
element of the physics programme, just as d-Au collisions are at
RHIC~\cite{SatogatadAu}.
A preliminary study~\cite{EPAC06pPb} has shown that the injector chains
for protons and ions can work in tandem to efficiently fill the two LHC
rings with matching bunch trains.
However the two-in-one magnet
design of the LHC (as opposed to the separate magnets of RHIC) means
that  provision
of hybrid collisions in the LHC gives rise to
quite different beam dynamics.
Concerns have been raised about different revolution frequencies
during injection and part of the energy ramp and the consequent moving
beam-beam encounters.
At present, \cite{EPAC06pPb} gives
plausible indications that an acceptable luminosity can be
obtained or even surpassed.

\item[A-A collisions] of lighter ions such as Ar, Ca, \ldots, the choice
of ion being determined by the physics requirements and ease of
production by the ion source.

%% \item[Pb-Pb collisions] at lower energies or, perhaps, somewhat higher luminosity.

%% \item[p-A collisions] with lighter nuclei.

\item[Electron-ion collisions:] If, one day, e$^{\pm}$-p collisions are
implemented (the LHeC option) then it would be natural to provide e$^{\pm}$-A
collisions also.

\end{description}

To widely varying degrees, each of these would require further study and
adaptations of the CERN accelerator chain and the LHC rings themselves.
 Detailed scheduling will have to take into account other uses and
upgrades to the LHC in the years to come.

\section{Conclusions}\label{sec:summary}

\begin{itemize}

\item The LHC is on track for the first proton beams and collisions in
summer 2008.  The schedule nevertheless remains tight.

\item The first nucleus-nucleus (Pb-Pb) collision  run is expected at
the end of 2009.  The timing of this is very sensitive to the
scheduling of beam time and resources for the ion injectors in 2009 and
to the allocation of LHC beam time.

\item The Pb-Pb luminosity is limited by new beam physics, particularly
nuclear electromagnetic interactions that lead to energy deposition in
superconducting magnet coils.  Measures are being taken to monitor and
alleviate these effects.  Furthermore, our understanding has been
steadily improving and subjected to experimental tests.

\item Integrated luminosity per experiment decreases with the number of
active experiments, particularly for smaller $\bstar$.

\item The programme for collision species beyond the baseline p-p and
Pb-Pb remains to be established and studied.

\end{itemize}

\subsubsection*{Acknowledgements}

This paper outlined the work of many people, over many years.  I
particularly thank G.~Bellodi, H.~Braun, R.~Bruce, C.~Carli, L.~Evans,
A.~Ferrari, W.~Fischer, S.~Gilardoni, D.~Kuchler, D.~Manglunki, S.~Maury
and G.~Smirnov for specific input and discussions.

\section*{References}

\bibliography{QM2008Jowett}

\end{document}